\documentclass{aa}
\usepackage{graphicx}
\usepackage{natbib}
\usepackage{amssymb}

\begin{document}

\title{G181.1+9.5, a new high-latitude low-surface brightness SNR}

\author{Roland Kothes \inst{1} 
\and Patricia Reich \inst{2}
\and Tyler J. Foster \inst{3}
\and Wolfgang Reich \inst{2}
}

\titlerunning{New SNR G181.1+9.5}
\authorrunning{Kothes et al.}


\institute{National Research Council Canada,
              Herzberg Programs in Astronomy and Astrophysics,
              Dominion Radio Astrophysical Observatory,
              P.O. Box 248, Penticton, British Columbia, 
              V2A 6J9, Canada
\and
              Max-Planck-Institut f\"ur Radioastronomie,
	      Auf dem H\"ugel 69,
	      D-53121 Bonn,
	      Germany
\and
              Department of Physics and Astronomy, Brandon University, 
              270 18th Street, Brandon, MB R7A 6A9, Canada
}

\date{Received; accepted}
  \abstract 
   {More than $90~\%$ of the known Milky Way supernova remnants are within $5\degr$ of the Galactic Plane. 
   The discovery of the new high-latitude SNR G181.1+9.5 will give us the opportunity to learn more about the 
   environment 
   and magnetic field at the interface between disk and halo of our Galaxy.}
   {We present the discovery of the supernova remnant G181.1+9.5, a new high-latitude SNR, serendipitously discovered in 
   an ongoing survey of the Galactic Anti-centre High-Velocity Cloud complex, observed with the DRAO Synthesis 
   Telescope in the 21~cm radio continuum and \ion{H}{i} spectral line.
   }
   {We use radio continuum observations (including the linearly polarized component) at 
   1420~MHz (observed with the DRAO ST) and 4850~MHz (observed with the 
   Effelsberg 100-m radio telescope) to map G181.1+9.5 and determine its nature as a SNR.
   High-resolution 21~cm \ion{H}{i} line observations and \ion{H}{i} emission and absorption 
   spectra reveal the physical characteristics of its local interstellar environment. 
   Finally, we estimate the 
   basic physical parameters of G181.1+9.5 using models for highly-evolved SNRs.}
   {G181.1+9.5 has a circular shell-like morphology with a radius of about 16~pc at a distance
   of 1.5~kpc some 250~pc above the mid-plane. The radio observations reveal highly linearly
   polarized emission with a non-thermal spectrum.
   Archival ROSAT X-ray data reveal high-energy emission from the interior of G181.1+9.5
   indicative of the presence of shock-heated ejecta. The SNR is in the advanced radiative phase
   of SNR evolution, expanding into the HVC inter-cloud medium with a density of 
   $n_{\textrm{\tiny HI}}\approx 1$~cm$^{-3}$. 
      Basic physical attributes of G181.1+9.5 calculated 
   with radiative SNR models show an upper-limit age of 16,000 years, a swept-up mass of more than
   $300$~M$_{\odot}$, and an ambient density in agreement with that estimated from 
   \ion{H}{i} observations.
      }
   {G181.1+9.5 shows all characteristics of a typical mature shell-type SNR, but its observed 
   faintness is unusual and requires further study.}

\maketitle

\section{Introduction}
The completion of the Canadian Galactic Plane Survey \citep[CGPS,][]{tayl03} has given the
community an incredibly broad view of large-scale dynamics of the Interstellar 
Medium (ISM) while preserving its spectacular details. Across the region of
the sky surveyed by the CGPS, the Galactic plane warps upward to large $z$-height 
distances, while the outer disk flares to several kpc in thickness. Because of 
its focus on the Galactic plane, the CGPS missed much of the off-plane ISM 
structure and activity. This was the motivation for a new smaller 
survey of the medium-latitude region towards the Galactic anti-centre
$\ell\sim$180$\degr$. 

The plethora of supernova remnants (SNRs), HII regions, high-velocity HI clouds 
(HVCs), and super-shells in this region is evidence of an extraordinarily 
energetic ISM. Two very large complexes of HVCs are seen in this area 
\citep[collectively known as the Anti-centre Chain of HVCs;][and references 
therein]{tama97}, as well as an \ion{H}{i} super-shell \citep[the {}``Anti-centre 
Shell'' or ACS;][]{heil84}, at least seven supernova remnants, and myriad
\ion{H}{ii} regions \citep[Sh2-232 to Sh2-237,][]{fost15}. To produce the ACS
alone would take the equivalent energy of 100 canonical supernova explosions 
\citep[10$^{53}$~erg;][]{kulk85}. All of this is happening in a comparatively small 
pie slice around the centre (3 kpc wide at R =10 kpc Galactocentric distance). 
The high level of activity in the plane here is very likely feeding energy to 
the ISM in the higher-latitude regions, and creating synergistic disk-halo
activity there that remains unexplored in high-resolution centimetre bands.

It is not necessarily surprising then that during the course of this survey we 
have discovered a new supernova remnant in this region. G181.1+9.5 
(R.A.(J2000) = $6^h 26^m 45.5^s$, DEC(J2000) = $+32\degr 31\arcmin$) is only
the latest in a long line of SNRs discovered with data from the DRAO Synthesis
Telescope \citep[DRAO ST,][]{land00}. Over the last 15 years, a total of 14 new supernova remnants 
and pulsar wind nebulae have been found in the Outer Milky Way Galaxy mostly 
with data from the Canadian Galactic Plane Survey
\citep{koth01b,koth03b,koth05,tian07,fost13,koth14,gerb14}.
SNR G181.1+9.5 is, 
however, no ordinary SNR. It has an unusual position at $\sim$10$\degr$ above the 
Galactic plane, is remarkably spherically-symmetric, and displays the maximum 
fractional polarization at 4850~MHz and 1420~MHz one would expect of synchrotron 
emission from a shock-compressed shell. On the sky, G181.1+9.5 lies towards the 
Northern filament of the HVC complex known as the Anti-centre Chain of HVCs 
\citep[][and references therein]{tama97}. In this paper, we present evidence 
that suggests the SNR is likely physically within the complex and interacting 
with an HVC. 

The Galactic plane ISM is a highly complex environment with stellar wind 
bubbles, chimneys, and twisted magnetic fields. Since evolution and morphology of 
SNRs is strongly
influenced by their environment, G181.1+9.5 perhaps represents the best opportunity yet to 
study SNR formation and dynamics in a relatively pristine and simplified 
environment. Further, this is the first example of a SNR evolving into the 
inter-cloud medium of an HVC complex. 

In this paper, we present the discovery of 
G181.1+9.5 and some of its more basic observed features and physical 
characteristics. The hope is to motivate and enable further new observational 
studies and modelling analyses by the community of this extraordinary 
supernova remnant.

\section{Observations and Data Processing}

\subsection{DRAO ST 1420~MHz Continuum Observations}
The radio continuum and linear polarization observations at 1420~MHz were obtained 
with the synthesis telescope at the Dominion Radio Astrophysical Observatory 
\citep{land00}, as part of a large survey of the Galactic anti-centre area above
the Galactic mid-plane. Observations of individual 
fields and data processing follow the procedures developed for the Canadian Galactic 
Plane Survey \citep{tayl03}. The survey area is covered by observations of individual fields
whose centres lie on a hexagonal grid of spacing $100\arcmin$. The data of individual fields
are carefully processed to remove artefacts and to obtain the highest possible dynamic range
using the routines developed by \citet{will99}.

\begin{figure}
   \centerline{\includegraphics[bb = 75 55 615 320,width=7.5cm,clip]{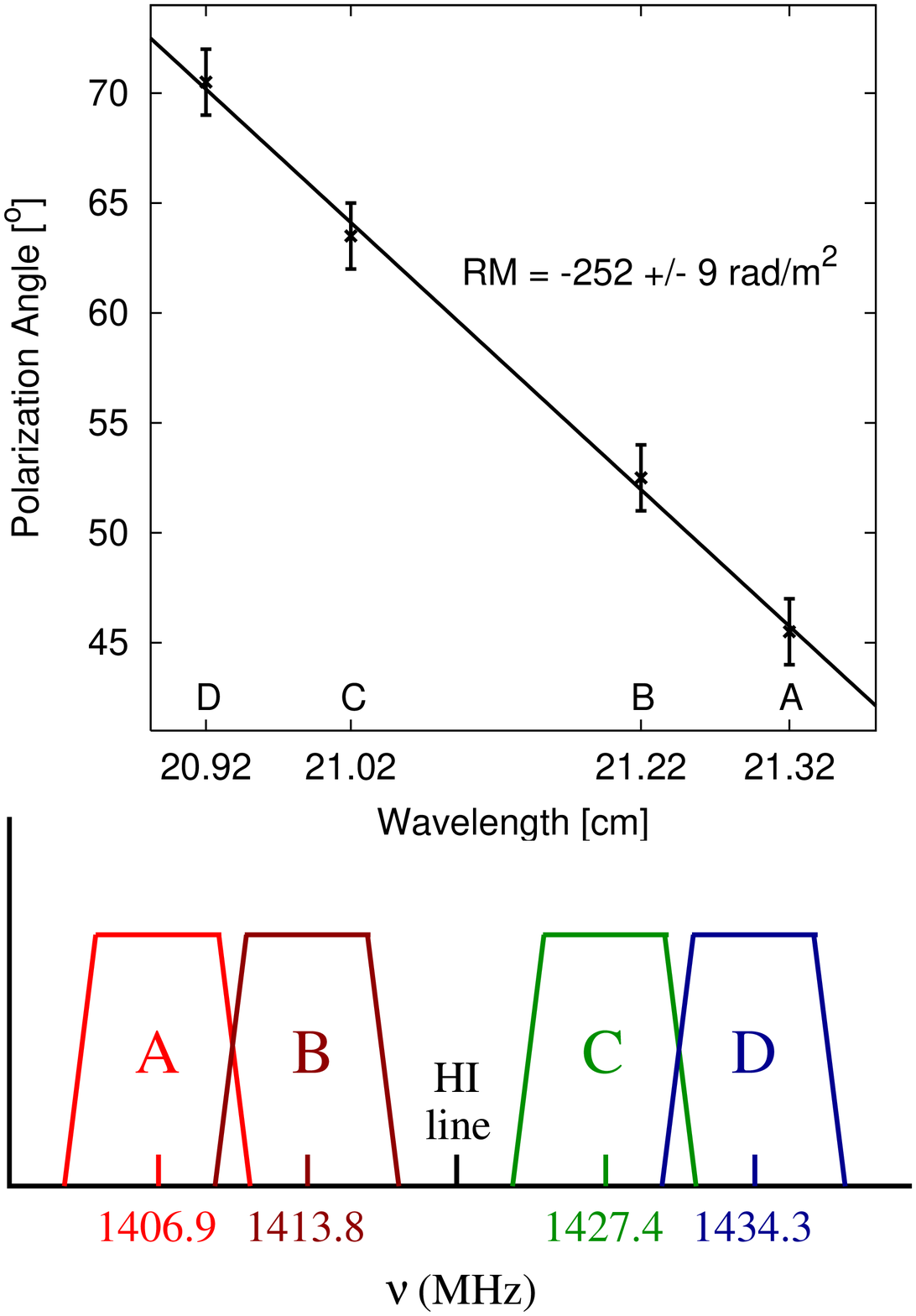}}
   \caption{Location of the four continuum bands relative to the HI band in DRAO ST observations.}
   \label{fig:abcd}
\end{figure}

The DRAO ST provides observations of the linearly polarized emission component in four separate 
bands of width 7.5~MHz at 1406.9~MHz, 1413.8~MHz, 1427.4~MHz, and 1434.3~MHz centred 
around the HI line to allow precise determination of rotation 
measures {\bf (see Figure~\ref{fig:abcd})}. Angular
resolution varies slightly across the final maps as cosec(Declination). At the centre
of G181.1+9.5, the resolution in the final radio continuum and polarization images is $95\arcsec \times 50\arcsec$
(DEC. $\times$ R.A.). The rms noise in the final combined maps is about 16\,mK or 
125\,$\mu$Jy\,beam$^{-1}$.

Accurate representation of all structures from the largest scales down to the resolution
limit is assured by incorporating absolutely calibrated linear polarization observations 
with the DRAO 26-m John A. Galt telescope 
\citep{woll06}, absolutely calibrated total-intensity observations with the Stockert 25-m telescope 
\citep{reic86}, 
and data of the 100-m Effelsberg radio telescope as
part of the Effelsberg Medium Latitude Survey \citep[EMLS,][]{emls}. Short spacing data
in Stokes $I$, $Q$, and $U$ are added to the interferometer images after suitable filtering in the Fourier domain
\citep{tayl03,land10}. At 1420~MHz in polarization,  short spacing data are added only to 
the combined Stokes $Q$ and $U$ maps not the individual bands.

\begin{figure}
   \centerline{\includegraphics[bb = 78 25 525 455,width=9cm,clip]{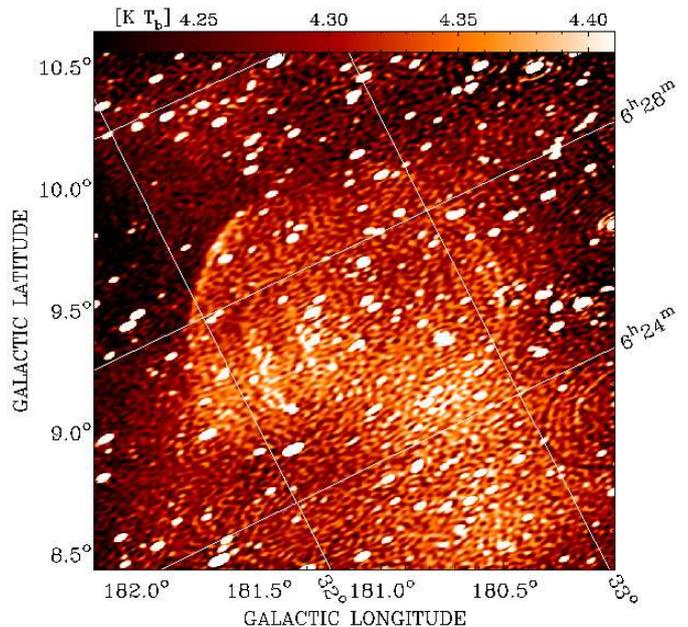}}
   \caption{Image of total intensity at 
   1420~MHz observed with the DRAO ST. Short spacings information observed with 
   the 100-m Effelsberg telescope and
   the 25-m Stockert radio telescope have been added in the Fourier domain (see
   Sect. 2.1 for more details). White lines indicate the coordinates in RA(J2000) and
   DEC(J2000) for convenience.}
   \label{fig:alli21}
\end{figure}

\begin{figure*}
   \begin{minipage}{8.9cm}
     \centerline{\includegraphics[bb = 75 25 482 450,height=9cm,clip]{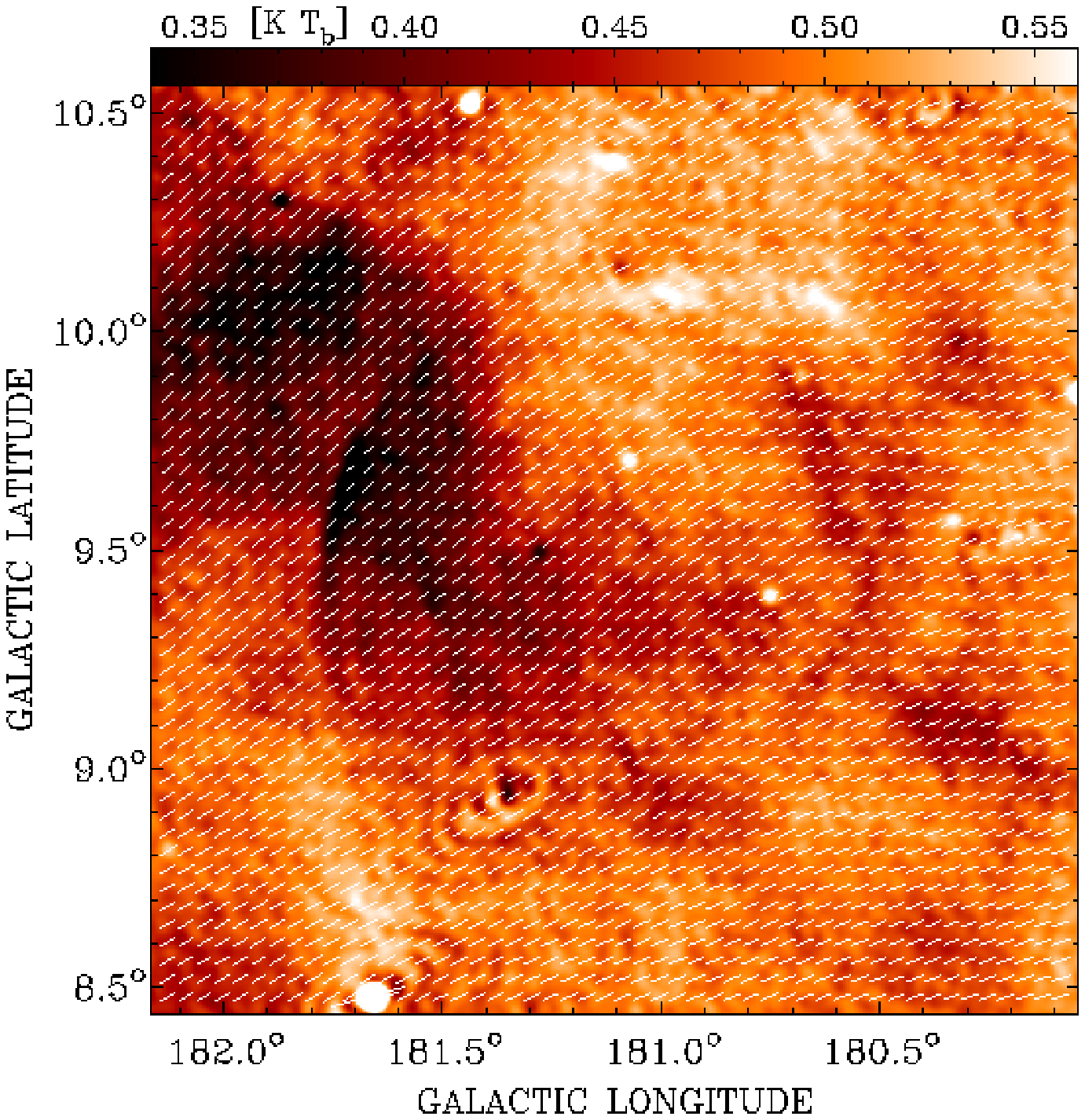}}
   \end{minipage}
   \begin{minipage}{8.9cm}
     \centerline{\includegraphics[bb = 74 54 502 486,height=9cm,clip]{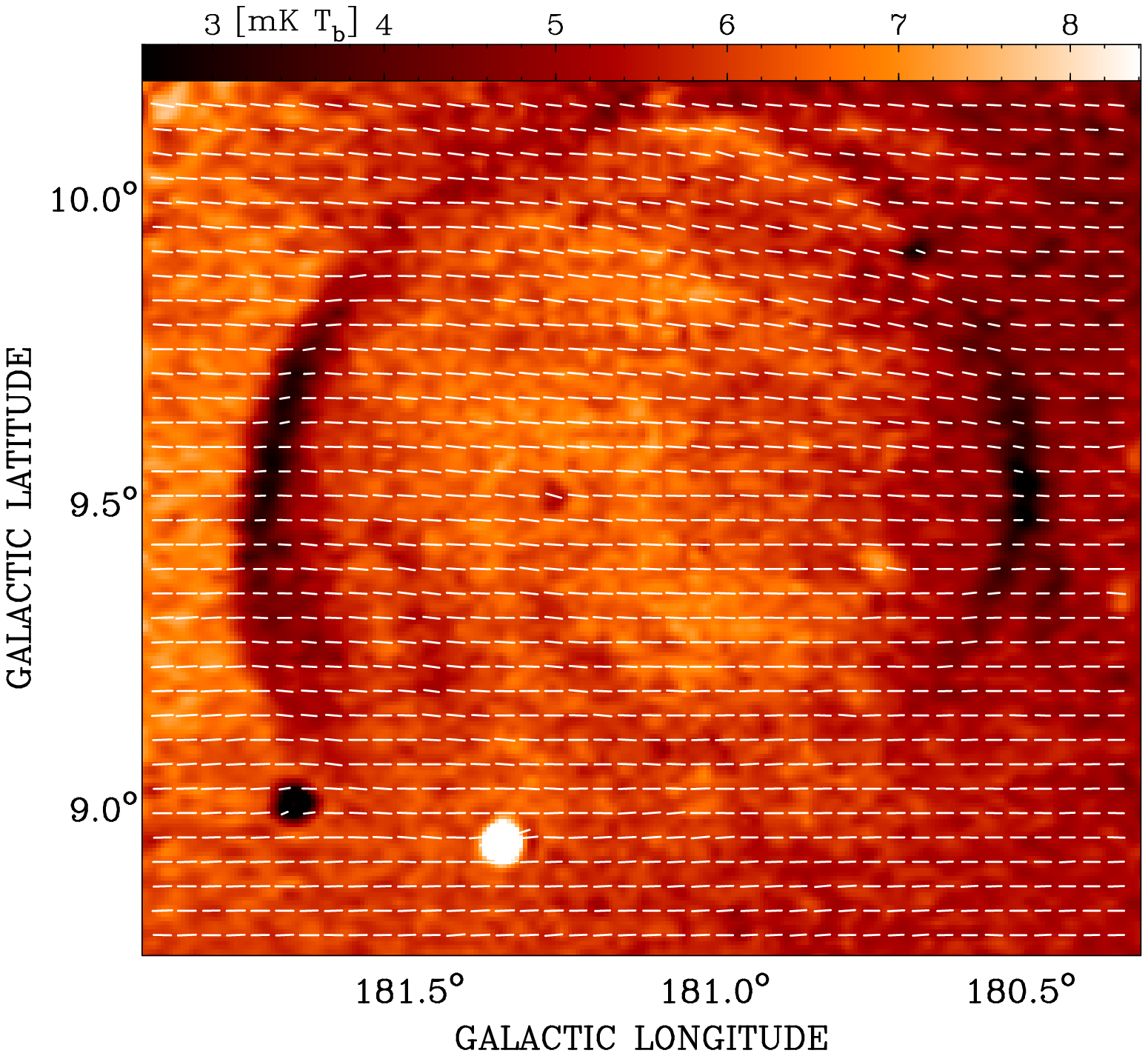}}
   \end{minipage}
   \caption{Images of polarized intensity at 1420~MHz observed with the DRAO ST ({\it left})
   and 4850~MHz observed with the Effelsberg telescope ({\it right}). Short spacings information 
   for 1420~MHz observed with the 100-m Effelsberg telescope and the DRAO 26-m John A. Galt 
   radio telescope and at 4850~MHz extrapolated from WMAP
   have been added in the Fourier domain (see
   section 2.1 and 2.2 for more details). The overlaid
   vectors in the polarized intensity images are shown in B-field direction.}
   \label{fig:allpi}
\end{figure*}

The final images in total intensity and polarized intensity at 1420~MHz are shown in Figs.~\ref{fig:alli21}
and \ref{fig:allpi}.

\subsection{Effelsberg Observations}

A second frequency is required to determine the characteristics of the radio continuum
spectrum of G181.1+9.5 to prove it is a SNR. We made 4850-MHz total-intensity and 
linear-polarization observations with 
the two-channel double-beam receiver installed in the secondary focus of the 
Effelsberg 100-m telescope. Between December 2013 and May 2015, a large number 
of maps was observed at night time and only at clear skies to reject solar
interference and weather effects. Due to strong 
unpolarized RFI of unknown origin, about half of the collected total-intensity 
data could not be used. All observation parameters are listed in Table~\ref{tab:eff}. 
Technical details including a receiver block diagram can be found at 
http://www.mpifr-bonn.mpg.de/effelsberg/astronomers under the item 
\textquoteleft receiver list / calibration parameters\textquoteright. To cover 
G181.1+9.5, we need to map a large area in the sky. The standard mapping procedure for 
Effelsberg 4850-MHz dual-beam observations, scanning along azimuth direction and 
subsequent restoration, works well for map-sizes up to about
$60\arcmin$. However, this method will lose emission on larger scales. Therefore we decided to 
conduct the observations in single-beam mode by moving the telescope along 
Galactic longitude and Galactic latitude. 

The data for both feeds were processed independently following the standard 
procedures for Effelsberg continuum observations, which were based on the NOD2 
format \citep{hasl74}. For each observed map, we flagged distorted areas and 
spiky interference and made baseline improvements by using the method of 
unsharp masking \citep{sofu79}. The maps were calibrated for each observing 
night in respect to the calibration sources 3C\,286 and 3C\,138, as listed in 
Table~\ref{tab:eff}.

\begin{table}[thp]
\caption{Effelsberg 4850~MHz Observational Parameters }
\label{tab:eff}
\centering
\begin{tabular}{lrrr}
\hline\hline
\\[-0.25cm]
\multicolumn{1}{c}{Data}  &\multicolumn{1}{c}{Effelsberg 4850~MHz}\\[0.1cm]
\hline             
\\[-0.25cm]
Frequency [MHz]                &4850  \\
Bandwidth [MHz]                &500   \\
HPBW[$\arcmin$]                &2.45   \\
Scanning velocity [$\arcmin$/min]     & 60 \\
Step interval [$\arcmin$]              & 1  \\
Size [$\degr$ $\times$ $\degr$]  & 1.7   $\times$ 1.5 \\
Number of complete maps ($I/U,Q$)   &16/30\\
Integration time ($I/U,Q$)[s/pixel]  &16/30   \\
rms ($I/U,Q$)[mK\ $T_b$]      &0.4/0.1         \\[0.1cm]
\hline
\\[-0.25cm]
Main calibrators                &3C\,286/3C\,138   \\
Flux densities [Jy]              &7.5/4.1     \\
Polarization percentages        &11.3\%/11.1\%  \\
Polarization angles [$\degr$]   &33/-11 \\[0.1cm]
\hline
\end{tabular}
\end{table}

To include the $I$, $Q$, and $U$ data from the second feed, which has an offset of 8\farcm2 in 
azimuth relative to the first one, we had to apply a special transformation, 
where we have to make sure that this process does not introduce smoothing or 
other distortions. For each map pixel from the first feed, which is in Galactic 
coordinates, we calculated 
its azimuth and elevation when observed and added the azimuth offset of the second
feed. We 
computed its exact position in Galactic coordinates, which usually 
deviated from the $1\arcmin$-grid of the map. Thus we transformed the data into a map 
with a $10\times$ finer grid ($6\arcsec$) and re-gridded stepwise to a 
$1\arcmin$-grid to be combined with the maps from the first feed.

The final map was obtained by combining the individually observed maps in the Fourier domain using the 
\textquotedblleft PLAIT\textquotedblright~algorithm described by \citet{emer88}.

The absolute level of polarized emission at 4850~MHz was
calculated following the method discussed by \citet{sun07}
by extrapolating the K-band WMAP (Wilkinson Microwave Anisotropy Probe) observations of Stokes $U$
and $Q$ from 22.8~GHz to 4850~MHz. We used the WMAP 9-year
data release \citep{benn13} and assumed
a temperature spectral index of $\beta$ = 3.2 ($T_{B} \sim \nu^{-\beta}$), which is
typical for this area relative to the 1420~MHz DRAO
polarization survey \citep{woll06}. At these high
frequencies, the influence of Galactic Faraday rotation
can be neglected. We convolved the $U$ and $Q$ maps at both
frequencies to 2$\degr$ and added the difference maps
to the 4850-MHz maps at full angular
resolution. The smoothing beam is uncritical to the result
and a different spectral index has almost no affect on
the morphology of the $U$ and $Q$ images, but the intensity
level. The resulting 4850-MHz polarized-intensity map is
shown in Fig. 2 together with the corresponding 1420-MHz
map at an absolute intensity level.

\begin{figure*}[tb]
   \centerline{\includegraphics[bb = 35 115 551 440,width=17.0cm,clip]{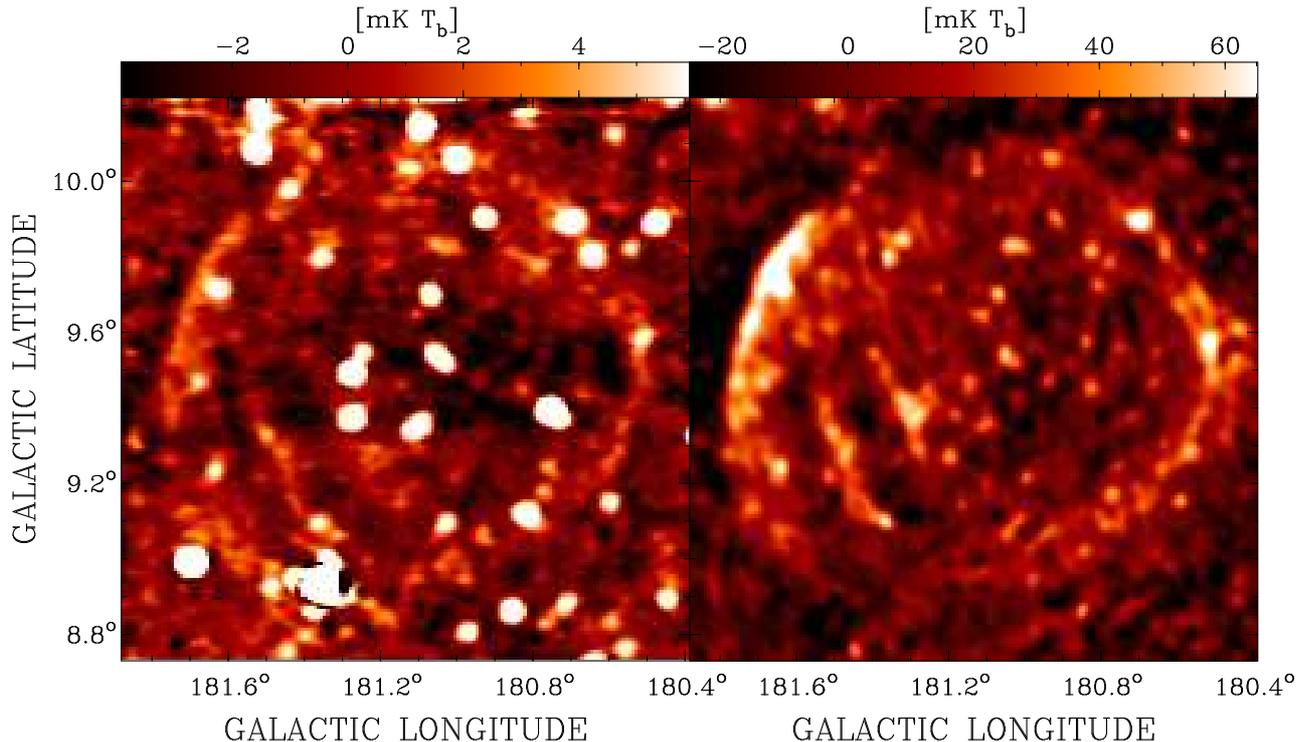}}
   \caption{Total power images of G181.1+9.5 at 4850~MHz ({\it left}) and 1420~MHz ({\it right}). In the
   1420-MHz map the brighter point sources have been removed and the map was convolved to
   the same resolution as the 4850-MHz map of $2\farcm45$.}
   \label{fig:tp}
\end{figure*}

\subsection{\ion{H}{i} Observations}
To explore the gaseous ISM environment within which G181.1+9.5 is found, we 
used 12 fields from the DRAO ST Anti-Centre survey in the 21~cm spectral line to characterize the
HI environment on and around the object. We used uniform weighting of the antenna elements to 
achieve a synthesized beam in this part of the sky approximately 
95$\arcsec\times$50$\arcsec$ (DEC. $\times$ R.A.). Data are calibrated to the brightness temperature 
scale with observations of sources 3C\,48, 3C\,147 and 3C\,295 as described in 
\citet{land00}, and a beam scale factor of 
$T_{B}(\textrm{K})/S(\textrm{mJy~beam}^{-1}) = 0.127$. The overlap of the 12 
fields lowers the noise in each velocity channel of the stacked mosaic,
resulting in a final sensitivity of $\Delta T_{B}=$1.3~K per 
0.824~km~s$^{-1}$-wide channel. Spectrally, data have velocity resolution of 
1.32~km~s$^{-1}$. 

While we use interferometer-only data to measure optical 
depths through the HVCs via \ion{H}{i} continuum absorption, we have also made 
a cube with all spatial frequencies from the largest scales down to the resolution
limit of the DRAO ST by incorporating single-dish \ion{H}{i} 
data as observed by the Effelsberg \ion{H}{i} Survey 
\citep[EBHIS;][; with spatial and velocity resolutions of $\sim 11\arcmin$ and 
1.4~km~s$^{-1}$, respectively]{wink16}. 
After smoothing to the same resolution as the 4850-MHz total-power data of $2\farcm45$, we inspected 
this data cube carefully for apparent associations between the radio continuum emission structure and 
the distribution of the neutral hydrogen gas (see Sec.~3.4).

\section{Results}

\subsection{G181.1+9.5 in Radio Continuum}

Maps of total intensity at 4850~MHz and 1420~MHz of the supernova remnant 
candidate G181.1+9.5 are shown in Figs.~\ref{fig:alli21} and \ref{fig:tp}. In 
Fig.~\ref{fig:alli21} the 1420-MHz image is shown at the original resolution of the 
DRAO ST with short spacing data from the EMLS included. G181.1+9.5 is
an extremely low-surface brightness almost circular object with thin shells on both sides.
It confuses with an emission plateau that extends from the top of the source to the south.

In Fig.~\ref{fig:tp}, more details of the overall structure are obvious. At both frequencies,
the candidate SNR is barely visible above the background emission, indicating that we are close to the
confusion limit. In addition to the two shells, two arc-like features are 
visible to the lower left on the inside. Their morphology strongly
suggests that they are related to G181.1+9.5.
The full diameter of G181.1+9.5 is about $74\arcmin$. For a more detailed discussion of the 
physical scale of G181.1+9.5 see Sect. 4.2.
Due to the low surface-brightness, the confusion with faint point-like sources, and a varying
background, it is not possible to determine reliable integrated flux densities at either 
frequency. But the most prominent part of the SNR, the Eastern shell, is about 90 - 100 mK
above background at 1420~MHz and about 4 - 5 mK at 4850~MHz. This indicates
a spectral index $\alpha$ ($S \sim \nu^{-\alpha}$, $S$: flux density, $\nu$: frequency)
between 0.3 and 0.6, typical for a shell-type SNR.

\begin{figure}[htb]
   \centerline{\includegraphics[bb = 13 13 526 416,width=8.5cm,clip]{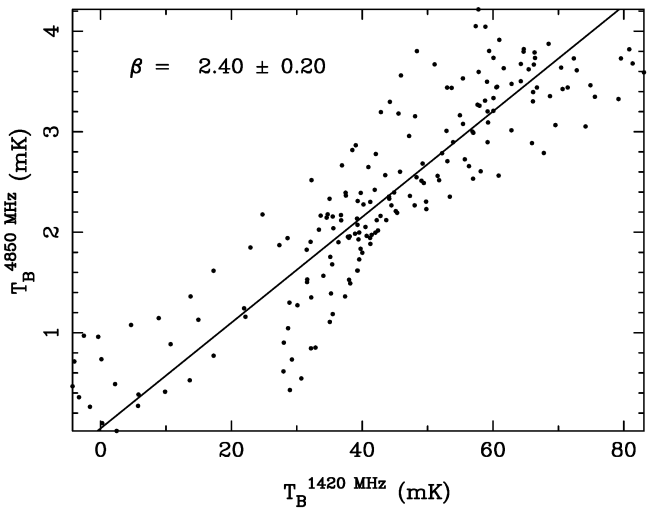}}
   \caption{TT-plot of the bright eastern shell of G181.1+9.5 between 4850~MHz and 1420~MHz.
   Point sources were removed and the maps were convolved to a common resolution of
   $4\arcmin$. The resulting temperature
   spectral index is 
   $\beta = 2.4\pm 0.2$.}
   \label{fig:tt}
\end{figure}

\begin{figure*}[htb]
   \centerline{\includegraphics[bb = 30 115 555 440,width=17cm,clip]{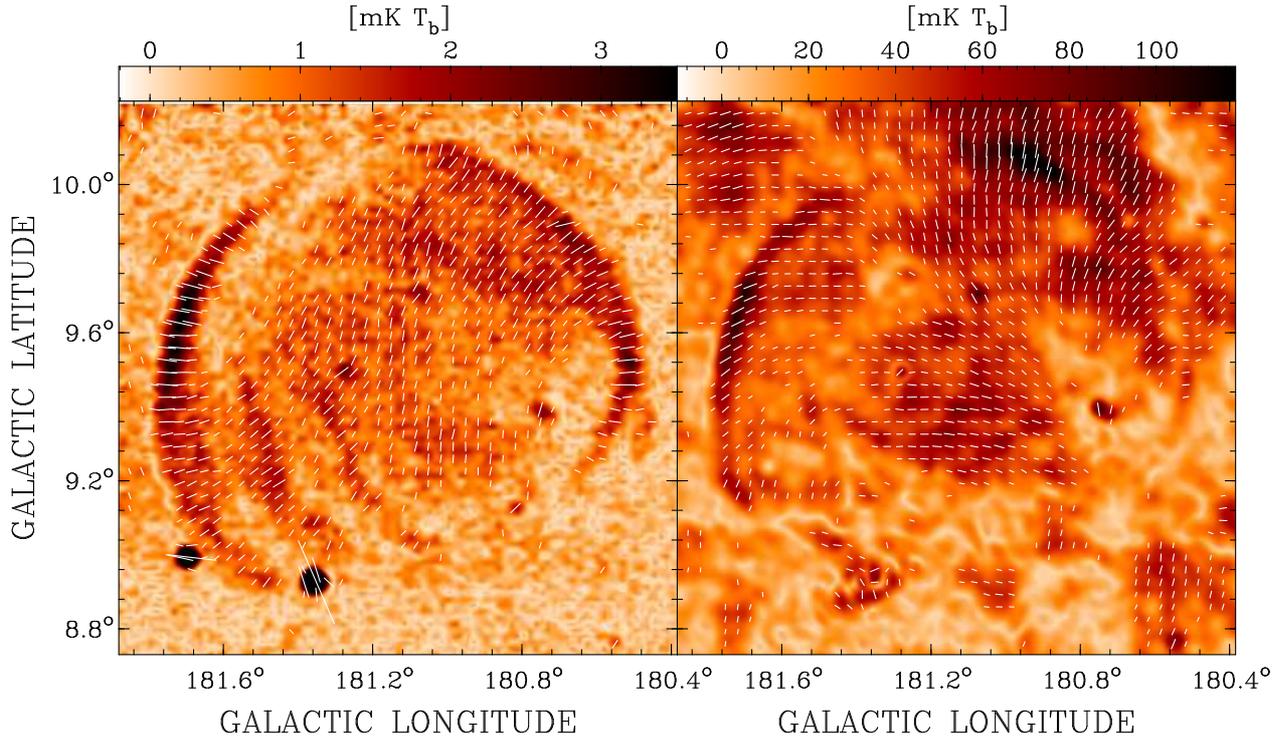}}
   \caption{Images of polarized intensity of G181.1+9.5 at 4850~MHz ({\it left}) and 1420~MHz 
   ({\it right}). The 1420-MHz maps were created from the DRAO ST observations alone and has 
   been convolved to the same resolution as the 4850-MHz map. 
   Vectors in observed E-field direction are overlaid. }
   \label{fig:pi}
\end{figure*}

We determined a more reliable spectral index by calculating ``TT-Plots'' over the more
prominent Eastern shell (Fig.~\ref{fig:tt}). This spectral-index determination method is 
described in detail in \citet{turt62}. TT-plots give 
reliable results for spectral indices of small-scale emission structures on-top of diffuse extended 
emission. The TT-plot (see Fig.~\ref{fig:tt}) results in a temperature spectral index of $\beta = 2.4\pm 0.2$.
This translates to a flux-density spectral index of 
$\alpha = 0.4\pm 0.2$ ($\beta = 2.0 + \alpha$) and confirms the non-thermal nature of this part of G181.1+9.5. 
However, both methods
produce results with large uncertainties.

\subsection{Linearly Polarized Emission from G181.1+9.5}

\begin{figure*}[ht]
   \centerline{\includegraphics[bb = 200 150 1210 735,width=17.5cm,clip]{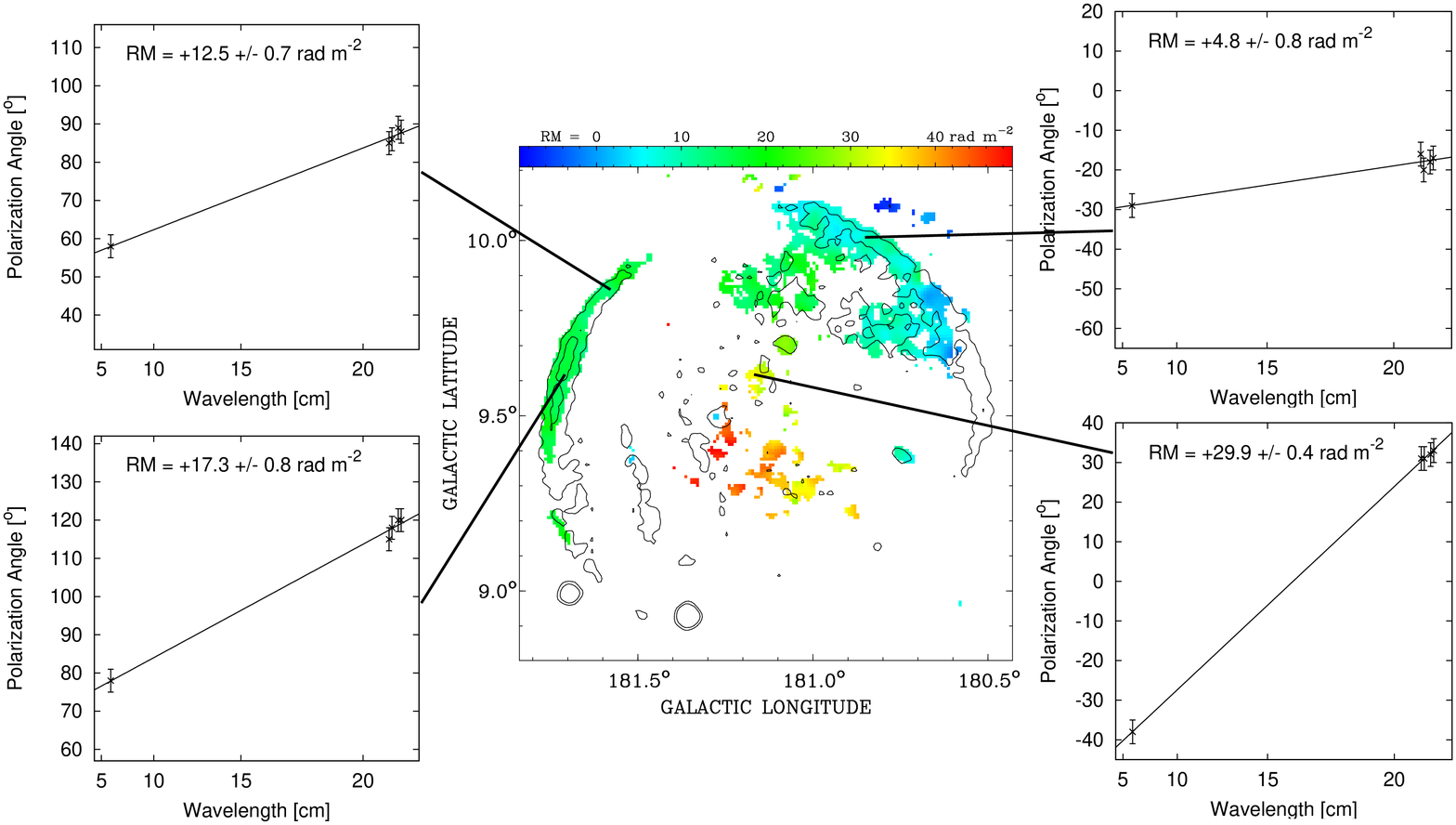}}
   \caption{RM map calculated between 4850~MHz and the four bands of the DRAO ST observations
   around 1420~MHz. Contours of the 4850-MHz $PI$ map are overlaid. Four sample RM spectra
   are shown representing the three different areas discussed in the text. The resulting RMs
   and their location on the map are indicated.}
   \label{fig:rmmap}
\end{figure*} 

Maps of polarized intensity ($PI$) with overlaid observed polarization vectors are shown in 
Figs.~\ref{fig:allpi} and \ref{fig:pi}. In 
Fig.~\ref{fig:allpi} the 1420-MHz and 4850-MHz $PI$ images are shown at the original resolution 
with short spacing data included. In all maps, the polarized
intensity was corrected for the so-called first order noise bias \citep[e.g.][]{ward74}:
\begin{equation}
PI = \sqrt{Q^2 + U^2 - (1.2 \sigma)^2}
\end{equation}
Here, $\sigma$ is the rms noise in the Q and U maps.

The polarized-intensity emission of G181.1+9.5 at 1420~MHz and 4850~MHz is seen ``in absorption'' relative
to the bright background polarization in this direction of our Galaxy (Figure~\ref{fig:allpi}). G181.1+9.5
seems to be located at the high-longitude outskirts of the so-called Fan region, a major 
feature of the linearly polarized
sky at radio frequencies that is visible above the Galactic plane in the second quadrant of
our Galaxy \citep[e.g. see Fig. 11 in][]{woll06}. 
In this direction of the Galaxy, the Fan region shows a surface brightness of about
0.5~K at 1420~MHz in polarized intensity (see Fig.~\ref{fig:allpi}) compared to less than 0.1~K for G181.1+9.5.
At 4850~MHz, the background is about 6~mK compared to maybe 2 - 3~mK for G181.1+9.5.
At both frequencies, the polarization angles must be quite different so that the polarized emission averages
out and gives the impression that the polarized signal of the SNR is absorbed. 

The maps shown in Fig.~\ref{fig:pi} do not contain large-scale emission information so that
most of the polarized intensity is seen in emission, in particular at 4850~MHz. At
1420~MHz, there are still polarization features outside of the remnant that seem to be 
unrelated. Those features could indicate fluctuations in the Fan region, that appear
on angular scales visible to the DRAO ST. Unfortunately, 
there are also polarization features projected inside G181.1+9.5, which could be
background features or those parts of the SNR that are either moving towards us or 
away from us. Therefore, a comparison of polarization angles
between 4850~MHz and 1420~MHz to determine rotation measures (RMs) is quite challenging.

The rotation measure $RM$ is defined by:
\begin{equation}
  \phi_{\rm obs} = \phi_0 + RM\,\lambda^2,
\end{equation}
where $\phi_{\rm obs}$ (rad) is the observed polarization angle at wavelength $\lambda$ (m) and
$\phi_0$ (rad) represents the intrinsic unrotated polarization angle. RM is related to the 
magnetic field $B_\parallel$ parallel to the line of sight,
the electron density $n_e$, and the path-length $dl$ by:
\begin{equation}
RM = 0.81 \int_l B_\parallel n_e dl.
\end{equation}

Because of the low signal-to-noise ratio in the total-power maps, we cannot determine precise
values for the percentage polarization. But a crude comparison of the surface brightness
of the eastern shell indicates a percentage polarization of 70 to 80~\% for the entire eastern
shell at 4850~MHz. At 1420~MHz, the upper part seems to be slightly lower polarized with
60 to 70~\% polarization while the lower part shows the same 70 to 80~\% we found at 4850~MHz.

At 4850~MHz, we can clearly see the two prominent shells and the two arc-like 
features inside the SNR (Figure~\ref{fig:pi}). There is also some diffuse emission in the interior. The
polarization vectors are radial relative to all shells and arcs, indicating tangential 
magnetic fields, which is expected for a mature shell-type supernova remnant. The integrated 
polarized emission at 4850~MHz is $150\pm 20$~mJy. Assuming a maximum possible fractional
polarization of 70~\% as the average for the synchrotron emission from the entire source
this indicates a lower limit of 210~mJy 
for the integrated total-power emission at 4850~MHz. Extrapolating this with a spectral index
of $\alpha=0.4$ to 1~GHz gives 400~mJy. With a full diameter of $74\arcmin$, this results in
the lowest 1~GHz radio surface brightness ever observed for a SNR of 
$\Sigma_{\rm 1 GHz} = 1.1\times 10^{-23}$~W\,m$^{-2}$\,Hz$^{-1}$\,sr$^{-1}$.

Despite the difficulties, we calculated a RM map between 4850~MHz and the four bands around 1420~MHz after
convolving the Stokes $Q$ and $U$ maps at the four DRAO ST bands to the same resolution as the 4850-MHz
observation of $2\farcm45$ (Fig.~\ref{fig:rmmap}). We added four sample RM diagrams to 
Fig.~\ref{fig:rmmap}
to demonstrate the quality of the results.

The Eastern shell of G181.1+9.5 is quite distinct from other polarization features at
both frequencies. Here we can easily compare the observed polarization angles at all five frequencies.
We derive an average RM over this entire shell of +15~rad\,m$^{-2}$ with
variations of $\pm 4$~rad\,m$^{-2}$. There is no distinct gradient or other tendency detectable 
over this shell.

In the diffuse emission at the centre of G181.1+9.5, the RMs are higher up to
a maximum of +40~rad\,m$^{-2}$ and on the Western shell, we find RMs between 0 and 
+5~rad\,m$^{-2}$. The latter values should be considered with reserve, since we
cannot estimate the influence of the background polarization visible outside and
maybe superimposed on the Western shell. Again, no gradients or any other tendencies are observed.
None of these values are significant for the 4850-MHz
observations, even an RM of +40~rad\,m$^{-2}$ would rotate the polarization angle by
less than $10^\circ$.

\begin{figure}[ht]
   \centerline{\includegraphics[bb = 60 115 555 600,width=8.5cm,clip]{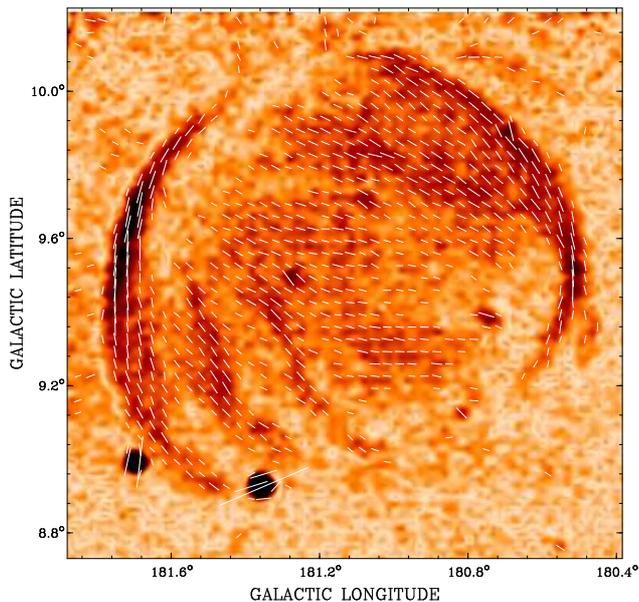}}
   \caption{Image of linearly polarized intensity observed at 4850~MHz with the 100-m Effelsberg
   radio telescope. Vectors in B-field direction are overlaid.}
   \label{fig:pi+b}
\end{figure}

Therefore, we neglect the affects of Faraday rotation on the polarization angle observed
at 4850~MHz and present the magnetic field direction projected to the plane of the sky 
by rotating the observed E-vectors by $90^\circ$. The resulting map is shown in 
Fig.~\ref{fig:pi+b}. The derived magnetic field configuration is tangential to the outer
shells and the two arc-like filaments in the interior, as expected for mature shell-type
supernova remnants. The diffuse interior emission shows magnetic fields that are overall 
parallel to the shells as expected from simulations \citep[e.g.][]{koth09,west16}. This is
a strong indicator of the magnetic field direction around G181.1+9.5.

\subsection{Detection in the ROSAT All-Sky-Survey}

\begin{figure}[!ht]
   \centerline{\includegraphics[bb = 60 115 555 600,width=8.5cm,clip]{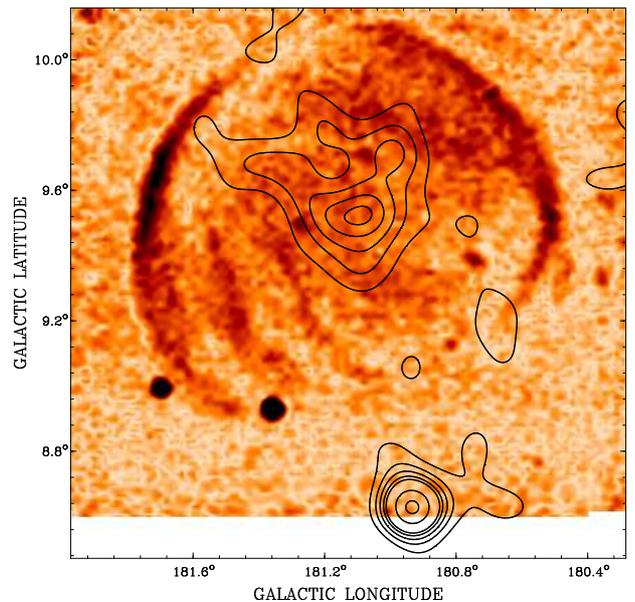}}
   \caption{4850-MHz polarized intensity image with X-ray contours from the RASS
   hard energy band overlaid. The RASS data have been convolved to a resolution of
   $10\arcmin$.}
   \label{fig:rass}
\end{figure}

We made a literature and archival search to find signatures of this SNR candidate
at other wavelength. We were successful when we inspected data from the ROSAT
All-Sky Survey database \citep[RASS,][]{voge99}. G181.1+9.5 is detectable in
the RASS hard-energy band (0.5 to 2.0~keV). In the soft band, the noise is probably 
too high and some of the emission might be absorbed by the foreground. In 
Fig.~\ref{fig:rass}, we display the polarized-intensity image with X-ray
contours overlaid. As a comparison, we left a known compact source in the field of
view of the image. 1RXS~J062246.2+321853 is an unidentified compact X-ray source from
the RASS \citep{voge99} with about 30 counts in the 0.5 to 2.0~keV energy range.
It can be seen at the bottom edge of the image in Fig.~\ref{fig:rass}.

The peak of the X-ray emission is exactly at the geometrical centre of G181.1+9.5
at a signal-to-noise ratio of about 12\,$\sigma$ in Fig.~\ref{fig:rass}.
Even after the convolution to $10\arcmin$ the emission shows a lot of substructure
indicating diffuse extended emission filling a significant part of the SNR
candidate's interior. This is very similar to the discovery of the CGPS SNRs
G85.4$+$0.7 and G85.9$-$0.6 \citep{koth01b}, which also displayed faint X-ray
emission in the high-energy band of the RASS. Follow-up observations of this
X-ray signature with XMM and CHANDRA found diffuse X-ray emission from the 
shock-heated supernova ejecta \citep{jack08}. The chemical composition of the ejecta from
G85.9$-$0.6 identified
this SNR as the remnant of a type Ia supernova explosion.

\subsection{The HI Environment of G181.1+9.5}
The majority of 21~cm line emission from neutral hydrogen in the field towards 
G181.1+9.5 is found in two distinct velocity regimes: disk and halo gas 
(+15~km~s$^{-1}\geq v_{\textrm{\tiny LSR}}\geq-$25~km~s$^{-1}$), and the realm 
of the high-velocity clouds in the region 
($-$75~km~s$^{-1}\geq v_{\textrm{\tiny LSR}}\geq-$95~km~s$^{-1}$). Because of 
the compression of long path-lengths into small velocity intervals inherent in 
anti-centre lines of sight, the best that we can say about the location of G181.1+9.5 
is either that it is found in the disk-halo and therefore is moving normally 
with circular Galactic rotation, or it is found among the HVCs (which are blind 
to Galactic rotation) and moving with them. Therefore, a search of the \ion{H}{i} 
data-cube for correspondence between the continuum shell and hydrogen emission 
features may lead us to one or the other location for G181.1+9.5.

\begin{figure*}[ht]
   \centerline{\includegraphics[bb = 30 120 530 735,width=16cm,clip]{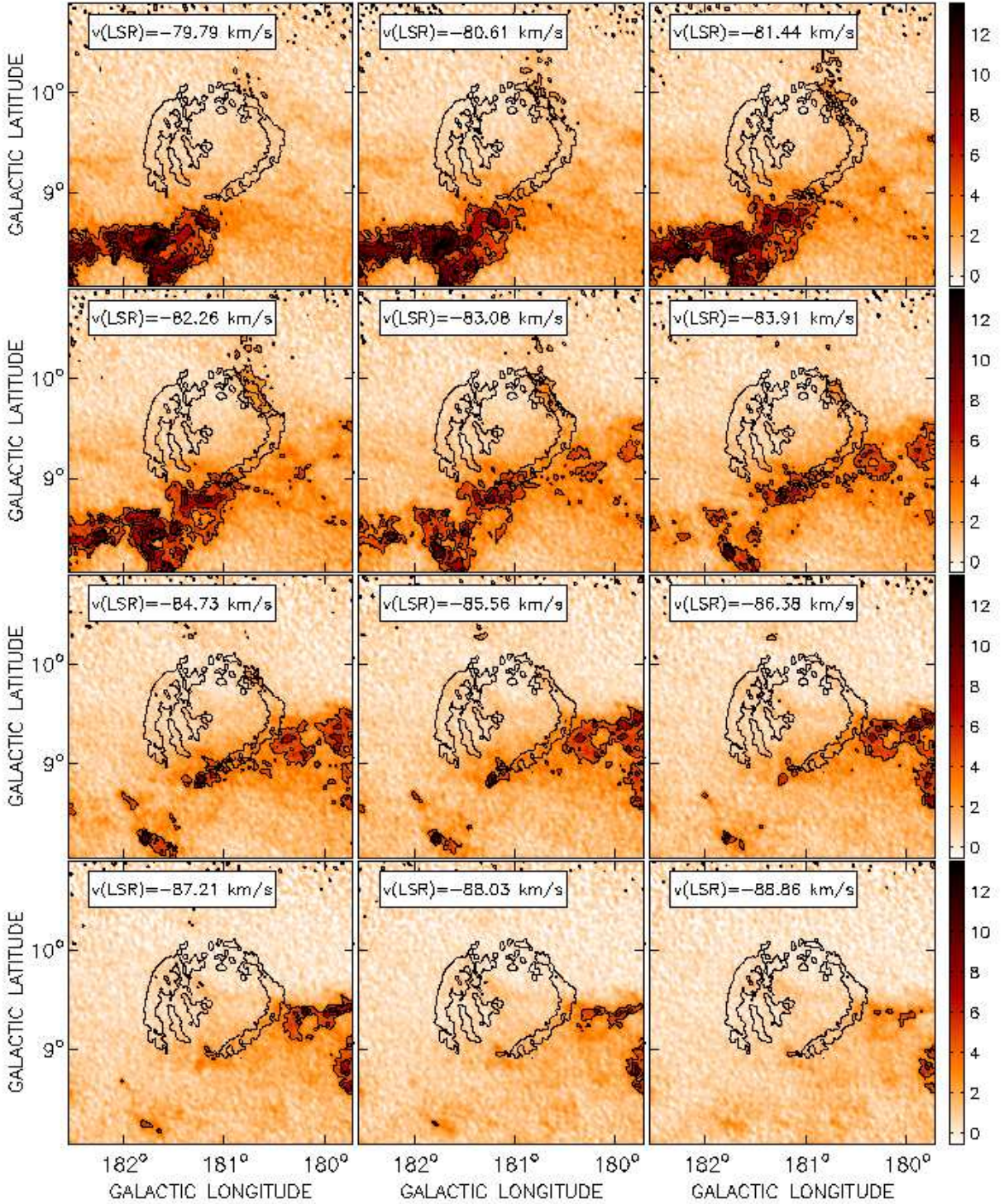}}
   \caption{Twelve 21~cm line-channel maps towards G181.1+9.5 in the velocity 
regime of the HVCs in the area. The colour-bar at right is 
K $T_{B}$. Contours around \ion{H}{i} emission 
clouds are at $T_B=$2,4,6,8~K, and a single thick black contour shows the 1420-MHz 
total-power appearance of the object.}
   \label{himaps}
\end{figure*}

\begin{figure}[ht]
   \centerline{\includegraphics[bb = 40 250 515 745,width=8.5cm,clip]{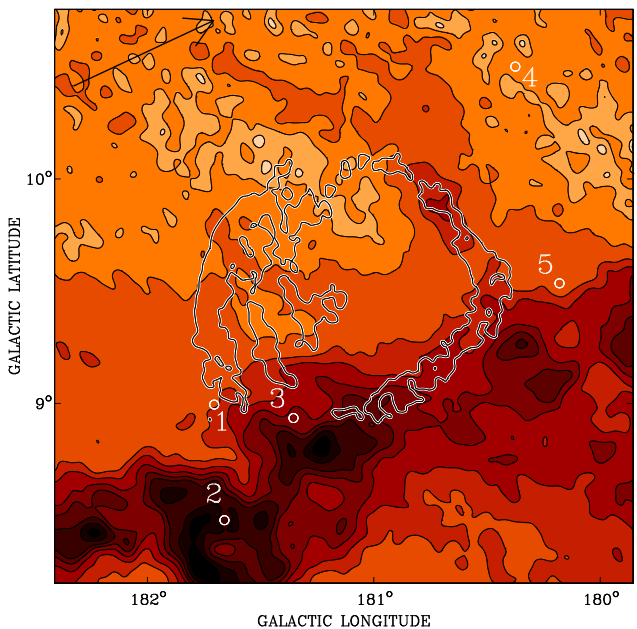}}
\caption{21~cm emission from neutral hydrogen, integrated over velocity range 
$-80 \geq v_{\textrm{\tiny LSR}}\geq-$86~km~s$^{-1}$ to highlight the 
correspondence of G181.1+9.5's shell with the HVCs in the 
Northern Anti-centre Shell. Bright \ion{H}{i} emission corresponds to dark-black 
shadings. The resolution of the map has been reduced to $4\arcmin$.
Positions of five 
unresolved continuum sources used to measure absorption spectra 
(Fig.~\ref{spectra}) are marked with white circles, with source numbers 1-5 
referred to in the text (Sec.~3.4). While the map is gridded in Galactic 
coordinates, for 
reference the arrow marked in the upper-left corner of the map points 
towards the North Celestial Pole.}
\label{inthi} 
\end{figure}

A search through the data-cube reveals no obvious correlation with the SNR and 
disk-halo \ion{H}{i} features. We find no cavities, no crescents of emission 
(such as swept-up gas by a stellar wind), no partial/complete bubbles enclosing 
the circular SNR shell, and no discrete isolated clouds appearing anywhere 
around the limb. This does not entirely rule out a connection with the disk 
and/or the halo gas. We do find a more convincing correlation between the shell 
and HVCs near $\sim -$80~km~s$^{-1}$. A detailed channel-by-channel view in 
Fig.~\ref{himaps} shows HVC clouds in emission, seen just on the outer edge 
of the Northwest shell, appearing at 1-to-2~o'clock in channels from 
$-$79.79~km~s$^{-1}$ to $-$84.73~km~s$^{-1}$. HVC clouds also appear at the
shell's Western edge at about 4-o'clock ($-$83.08~km~s$^{-1}$ to 
$-$88.03~km~s$^{-1}$), and around the projected Southern edge of the object 
(where no continuum shell is seen) from $-$80.61~km~s$^{-1}$ to about 
$-$85.56~km~s$^{-1}$. No \ion{H}{i} clouds associated with the HVC region 
appear inside the shell at any velocity. Figure~\ref{inthi} displays an 
integrated \ion{H}{i} map (averaged across the HVC velocities) showing the 
overall relationship of the shell with the clouds.    

Further evidence for a connection between HVC \ion{H}{i} and the object is 
found by tracing hydrogen in absorption at the shell's periphery. The optical 
depth $\tau$ and density $n_{\textsc{\tiny HI}}$~(cm$^{-3}$) of the HVC gas in 
the environment around G181.1+9.5 can be estimated from \ion{H}{i} continuum 
absorption lines towards unresolved continuum sources. Five appropriately 
bright sources are marked in Fig.~\ref{inthi} with white circles, and range in 
relative peak continuum brightness temperature (above background) from 
left-to-right: 22~K, 203~K, 188~K, 80~K, and 86~K (sources 1-5 respectively).

\begin{figure}[ht]
   \centerline{\includegraphics[bb = 25 265 570 705,width=8.5cm,clip]{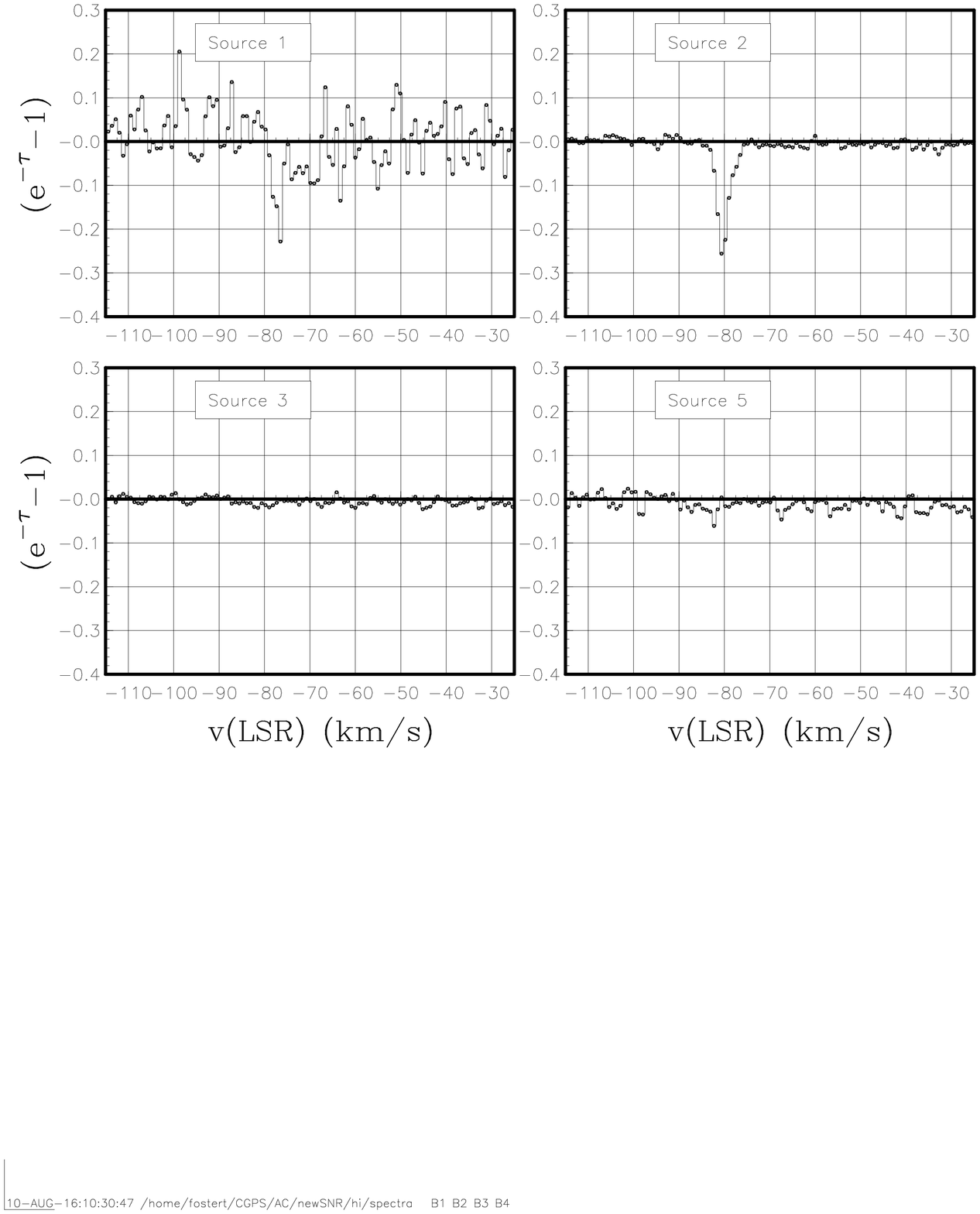}}
\caption{Absorption spectra towards four unresolved sources used to probe the 
ISM environment towards lines of sight near and off the shell of G181.1+9.5.}
\label{spectra} 
\end{figure}

Absorption spectra for sources 1, 2, 3, and 5 are shown in Fig.~\ref{spectra}. Only
the range of velocities appropriate for the HVCs are plotted. We plot the 
dimensionless quantity:
\begin{equation}
e^{-\tau(v_{\textrm{\tiny LSR}})}-1 = 
\frac{T_{\textrm{on}}(v_{\textrm{\tiny LSR}})-T_{\textrm{off}}(v_{\textrm{\tiny
LSR}})}{T_{\textrm{cont}}}.
\end{equation}
Here $v_{\textrm{\tiny LSR}}$ is the radial velocity (relative to the Local 
Standard of Rest), $T_{\textrm{on}}$ is the brightness temperature observed at 
the continuum source's position, and $T_{\textrm{off}}$ is the average around 
the continuum source. $T_{\textrm{cont}}$ is the peak brightness temperature of 
the background continuum source measured at the same position as 
$T_{\textrm{on}}$.

Source~1 is serendipitously found just on the Eastern edge of the shell. Even 
though it has the highest rms noise, the spectrum towards this source shows a 
line with a peak optical depth $\tau\sim$0.260$\pm$0.064 at 
$v_{\textrm{\tiny LSR}}=-$76.5~km~s$^{-1}$, and is similar to that of the 
bright Source~2 ($\tau=$0.296$\pm$0.010), which is seen absorbed by a HVC in 
emission half a degree south of G181.1+9.5. The bright Source~3 is just off the
projected Southern edge of the object, where no continuum shell is seen, 
and shows essentially no absorption at HVC velocities. Finally the two other  
Sources~4 and 5 which are well west of G181.1+9.5 and in the inter-cloud region 
of the HVC complex also show no significant absorption lines at the velocities 
of the HVCs (only the spectrum towards Source~5 is shown in 
Fig.~\ref{spectra} to demonstrate this).

Thus, there appears to be cold \ion{H}{i} immediately off the shell's Eastern
edge (the location of Source~1). In addition the remnant seems to be cut off at 
the bottom where the HVC clouds are at their brightest, indicating
it could be limiting the expansion or breaking up the remnant there.
This and the correspondence of \ion{H}{i} emission with 
the shell of G181.1+9.5 leads us to conclude that this object is interacting 
with the HVC complex in the area. 

\section{Discussion}

The newly discovered radio source G181.1+9.5 shows the typical radio
morphology of a shell-type SNR. We found strong indication 
that the spectrum of the Eastern shell is non-thermal. The entire source
G181.1+9.5 is highly linearly polarized with the Eastern shell showing
a percentage polarization at the maximum expected for synchrotron emission.
In fact, extrapolating the fractional polarization we found in the Eastern 
shell to the rest of the object, makes this SNR the highest polarized 
shell-type SNR known, closely followed by G107.5$-$1.5 
\citep{koth03b}, G156.2+5.7 \citep{reic92,xu07}, and G182.4+4.3 \citep{koth98}, 
a highly polarized supernova remnant just about $5\degr$ below G181.1+9.5 in
the Galactic plane. The intrinsic polarization angles indicate a tangential 
magnetic field within the shell and the interior filaments of G181.1+9.5. We 
also found faint diffuse X-ray emission in the RASS energy band between 
0.5 and 2.0~keV, indicative of shocked heated ejecta. 
Therefore, we conclude that 
G181.1+9.5 is a mature shell-type supernova remnant.

The two polarized filaments inside the outside shells are likely related 
features, indicative of a highly structured environment. The two
filaments are visible in total power at both frequencies and in polarization
at 4850~MHz. They appear to be quite thin and contain a tangential magnetic field, 
therefore they must be highly compressed
and moving mostly sideways relative to the line of sight. Viewed from the
geometric centre, they expand towards the Southern part of the Eastern shell
which shows a lower surface brightness than the Northern part. Therefore these
two filaments are likely features inside the two major shells, which mark the
impact of the shockwave on small clouds. However, we did not find any evidence
of HI emission from clouds inside the SNR.

\subsection{Distance and Environment of G181.1+9.5} 

The magnetic field inside the SNR G181.1+9.5 is tangential to all of its shells 
and filaments and shows in
the centre an orientation that is mostly parallel to the direction of the shells
(see Fig.~\ref{fig:pi+b}), which indicates a magnetic field with an angle of about 
$60\degr$ with the Galactic plane. According
to the simulations of \citet{koth09}, this indicates that the SNR is expanding into an
ambient magnetic field with the same orientation. The Fan region,
however, contains a magnetic field that is parallel to the Galactic plane,
which can be nicely seen in the images of the 
WMAP observations \citep{page07}, but also in our
4850-MHz polarized intensity map including WMAP-based large-scale correction (see Fig.~\ref{fig:allpi}). Even
at a frequency of 1420~MHz the Fan region is more than 30\,\% linearly polarized
(Figs.~\ref{fig:alli21} and \ref{fig:allpi}) indicating this synchrotron emission originates from a 
long line of sight. Therefore, G181.1+9.5 cannot be somewhere in the interstellar medium 
along the line of sight, but must be either very far away in the halo, which could have
a different magnetic field orientation, or related to a distinct feature that is not part of the
Fan region. A location in the halo is unlikely, because at a large distance the SNR would
be unreasonably large and the ambient density would be too low. This indicates a
correlation with the HVC complexes, since those could have magnetic fields independent of the
large-scale Galactic magnetic field. In fact, simulations by \citet{sant99} show that 
HVCs colliding with the Galactic disk
can significantly deform the magnetic field, which could even be perpendicular to the 
plane for some time.

We have shown in Sect.~3.4 that G181.1+9.5 is very likely related to HVC
gas that is part of the huge Anti -- Centre Shell identified by 
\citet{heil84}, which extends for about 30$\degr$ in the sky. \citet{kulk86} 
used Ca K absorption features of stars with known distances and found that
the Anti-Centre Shell must be more distant than 0.5 kpc. The maximum distance is 
2.5 kpc, or may be even a little larger depending on the abundance of Ca. We adopt
the centre of those limits, $d = $1.5~kpc, as the heliocentric distance to G181.1+9.5 
in order to quantitatively estimate some of its physical characteristics.

The density of the faint \ion{H}{i} emission at the location of Source~1 is 
found from:
\begin{equation}
\frac{n_{\textsc{\tiny HI}}}{\left(\textrm{cm}^{-3}\right)}=1.82\times10^{18}
\int\frac{T_{\textrm{B}}\left(v\right)}{\left(\textsc{K}\right)}
\left[\frac{\tau\left(v\right)}{\left(1-e^{-\tau\left(v\right)}\right)}\right]
~\frac{dv}{dr}
\label{eqn1}
\end{equation}
where $dv$ (km~s$^{-1}$) is the velocity interval that corresponds to a path 
interval along the line-of-sight of $dr$ (cm). Since we don't know this a priori 
for the HVC region, we approximate $dv/dr$ here as the ratio of the 
velocity-width of the absorption feature (approximately 3.3~km~s$^{-1}$) to the 
spatial width of the absorbing cloud on the sky. We estimate this spatial width 
$dr\lesssim$7-12~pc by eye from the average angular width of HVC emission 
features related to this area of the shell ($16\arcmin$ - $28\arcmin$), and the adopted 
distance of 1.5~kpc. The factor $\tau/\left(1-e^{-\tau}\right)$ corrects the 
column density for optically thick \ion{H}{i} gas not seen in emission. The 
density calculated from Eq.~\ref{eqn1} is $n_{\tiny \textsc{HI}}=$0.9-1.5~cm$^{-3}$. 
Since Source~1 is located on the sky ahead of the shock, we take this 
as the ambient density of the medium into which G181.1+9.5 evolves. The
spherical structure of G181.1+9.5 and the symmetry of the radio shells also
indicate that the ambient medium density of the SNR is not varying significantly.

\subsection{Characteristics of the SNR G181.1+9.5}

\begin{figure}[ht]
   \centerline{\includegraphics[bb = 73 185 542 625,width=8.5cm,clip]{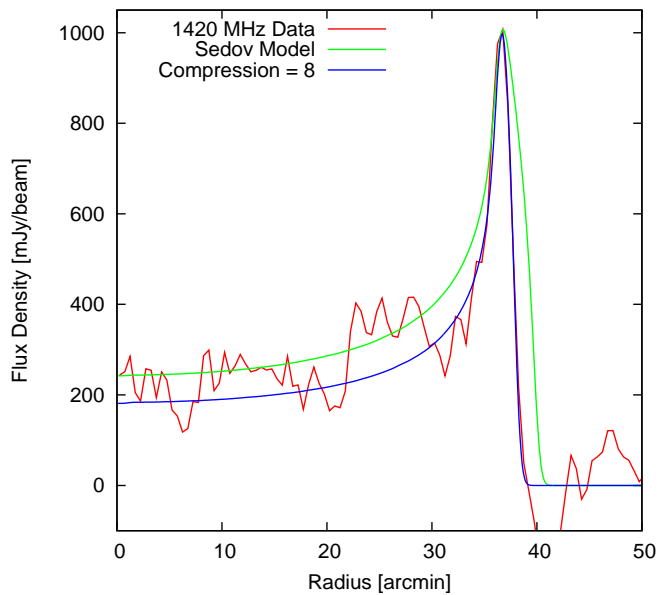}}
   \caption{Radial profile of the Eastern shell of G181.1+9.5, measured with the
   full resolution 1420~MHz DRAO ST only total power data (red). Simulated radial
   profiles in the Sedov phase (green) and a SNR with a compression ratio of 8 
   (blue) are shown for comparison.} 
   \label{fig:radprof}
\end{figure}

We produced a radial profile from the 1420-MHz total-power map at
full resolution, without the EMLS data added in, over the Eastern shell only. 
Compact sources have been removed from the map. The amplitudes of the profile are ring-averaged flux densities, centred at 
the geometric centre of G181.1+9.5 ($\ell = 181.13\degr$ and $b = 9.48\degr$). 
Individual rings are $30\arcsec$ wide.
This radial profile is shown in Fig.~\ref{fig:radprof}.

G181.1+9.5 has an angular radius of $37\arcmin$, which at a distance of 1.5~kpc
translates to a physical radius of about 16~pc. We simulated maps of SNRs 
with different compression ratios and calculated radial profiles after convolution to the
appropriate resolution. In those simulations, it has been assumed that magnetic field
strength and relativistic electron distribution are constant within the SNR's shell.
If the supernova remnant is in the so-called energy-preserving Sedov phase of its evolution, the 
compression ratio should be 4 \citep{sedo59}. It is obvious from
Fig.~\ref{fig:radprof} that the compression in the Eastern shell of G181.1+9.5 is
higher. We get a best agreement between model and observation for compression ratios 
of 8 and higher. At this point, the shell is apparently not resolved any more, which means
that the width of the shell in the profile is dominated by the resolution of our
observations. Nevertheless, it is obvious that G181.1+9.5 must be beyond the Sedov 
phase in the so-called radiative pressure-driven snowplow (PDS) phase, when SNRs become 
dominated by radiative energy losses.

We used the simulations by \citet{ciof88} to determine characteristics of G181.1+9.5
assuming it is in the transition between the Sedov and the PDS phase. At the beginning of
the PDS phase, \citet{ciof88} determined the following equations for the radius 
$R_{\tiny \textrm{PDS}}$, the velocity $v_{\tiny \textrm{PDS}}$, and the age of
the SNR $t_{\tiny \textrm{PDS}}$, assuming solar abundances:
\begin{equation}
  R_{\tiny \textrm{PDS}} = 14.0~\frac{E_{\tiny 51}^{2/7}}{n_{\tiny 0}^{3/7}}~~{\rm pc}
\end{equation}
\begin{equation}
  v_{\tiny \textrm{PDS}} = 413~E_{\tiny 51}^{1/14}~n_{\tiny 0}^{1/7}~~{\rm km\,s}^{-1}
\end{equation}
\begin{equation}
  t_{\tiny \textrm{PDS}} = 1.33\times 10^4~\frac{E_{\tiny 51}^{3/14}}{n_{\tiny 0}^{4/7}}~~{\rm yr}
\end{equation}
Here, $E_{\tiny 51}$ is the explosion energy of the supernova in $10^{51}$\,erg and $n_{\tiny 0}$
the ambient number density in cm$^{-3}$. Assuming that G181.1+9.5 has indeed entered the PDS phase
and the explosion energy is the canonical $10^{51}$\,erg, we calculate for the ambient number 
density a minimum of $n_{\tiny 0} \ge 0.7$\,cm$^{-3}$, for the
age a maximum of $t \le 16,000$\,yr, and for the current expansion velocity a minimum of
$v \ge 390$\,km\,s$^{-1}$. The mass of the material swept up by the supernova shock wave would be 
$M_{\tiny \rm sw} \ge 300$\,M$_\odot$. The constrains on the average ambient density very
well agree with the estimates for the density in the surrounding HVCs.

Although all characteristics estimated are typical values that would be expected from a normal
mature shell-type supernova remnant, it is not clear why this SNR is so faint. 

\section{Conclusions}

This paper presents radio observations and analysis of an uncatalogued object
found in the high-latitude region of the Galactic plane. New data at 1420~MHz and
4850~MHz show the object as a highly
symmetric circular shell with a radio spectrum indicative of non-thermal
emission. The shell is significantly polarized at radio frequencies up to   
4850~MHz, and has diffuse X-ray emission situated in its precise geometric  
centre. Based on these observed characteristics, we classify G181.1+9.5 as a
new supernova remnant. Radio polarizaton maps at two frequencies demonstrates
that the SNR has an intrinsic magnetic field that is
tangential to its shell, and modelling of the shell's thin edge shows this
SNR has evolved beyond the Sedov phase into the radiative (or {}``snowplow'')
phase. 

Based on its morphological association with hydrogen clouds and
diffuse emission in the Anticentre Chain of HVCs we suggest G181.1+9.5 is at
the same distance as the HVC complex, about 1.5~kpc. The diameter of the
shell is thus $\sim$32~pc.
Its location far away from any star formation activity, high 
above the Gaactic plane, indicates a supernova of type Ia and therefore a
White Dwarf as the progenitor star that travelled far away from its place of birth
before it exploded.

Aside from its normal characteristics for a SNR, G181.1+9.5 is a
remarkable object. It is
the faintest SNR yet observed, with a radio surface brightness at 1 GHz of
$\Sigma_{\rm 1\,GHz} \simeq$1.1$\times$10$^{-23}$~W~m$^{-2}$~Hz$^{-1}$~sr$^{-1}$, a factor of
three below the previous faintest known SNRs of \citet{fost13}.
G181.1+9.5 presents a rare opportunity to
study SNR evolution and offshoot topics (such as reverse-shock heating of explosion
ejecta, suggested by the X-rays in the object's centre) in a uniquely pristine
environment that is much different from the Galactic plane ISM.
Certainly, more follow-up observations of G181.1+9.5 are needed to confirm its
remarkable attributes, particularly in optical and X-ray bands, and we hope  
this discovery paper inspires and facilitates new observations by the ISM    
community of this unique object.

\begin{acknowledgements}
We wish to thank Dr. Tom Landecker (DRAO) and Prof. Ernst F\"urst (MPIfR) for careful reading of the manuscript. 
Special thanks go to the Effelsberg station manager, Dr. Alexander Kraus,
for repeated scheduling of the difficult 4850-MHz observation at best weather
conditions. We thank Dr. J\"urgen Kerp for helpful discussions on HVCs. We also 
thank Dr. Andrew Gray of DRAO for scheduling two additional DRAO synthesis 
fields to map the SNR shell with more sensitivity after its initial discovery. 
The Dominion Radio Astrophysical Observatory is a National Facility operated by 
the National Research Council Canada. We have made use of the ROSAT Data 
Archive of the Max-Planck-Institut f\"ur extraterrestrische Physik (MPE) at 
Garching, Germany. The data are partly based on observations with the 100-m telescope of 
the MPIfR (Max-Planck-Institut f\"ur Radioastronomie) at Effelsberg.
\end{acknowledgements}

\appendix

\bibliographystyle{aa}
\bibliography{kothes}

\end{document}